\documentclass[letterpaper,twocolumn,superscriptaddress,floatfix]{revtex4}
\usepackage[latin1]{inputenc}
\usepackage{bm}
\usepackage{multirow,amssymb,amsbsy,amsmath}
\usepackage{graphicx}
\usepackage{verbatim}
\makeatletter
\usepackage{pifont}
\makeatother

\begin{document}

\title{Revisiting Bohr's principle of complementarity using a quantum device}

\author{Jian-Shun Tang}
\affiliation{Key Laboratory of Quantum Information, University of
Science and Technology of China, CAS, Hefei, 230026, China}

\author{Yu-Long Li}
\affiliation{Key Laboratory of Quantum Information, University of
Science and Technology of China, CAS, Hefei, 230026, China}

\author{Chuan-Feng Li$\footnote{email:cfli@ustc.edu.cn}$}
\affiliation{Key Laboratory of Quantum Information, University of
Science and Technology of China, CAS, Hefei, 230026, China}

\author{Guang-Can Guo}
\affiliation{Key Laboratory of Quantum Information, University of
Science and Technology of China, CAS, Hefei, 230026, China}

\date{\today}

\begin{abstract}
Bohr's principle of complementarity lies at the central place of
quantum mechanics, according to which the light is chosen to behave
as a wave or particles, depending on some exclusive detecting
devices. Later, intermediate cases are found, but the total
information of the wave-like and particle-like behaviors are limited
by some inequalities. One of them is Englert-Greenberger (EG)
duality relation. This relation has been demonstrated by many
experiments with the classical detecting devices. Here by
introducing a quantum detecting device into the experiment, we find
the limit of the duality relation is exceeded due to the
interference between the photon's wave and particle properties.
However, our further results show that this experiment still obey a
generalized EG duality relation. The introducing of the quantum
device causes the new phenomenon, provides an generalization of the
complementarity principle, and opens new insights into our
understanding of quantum mechanics.
\end{abstract}

\maketitle

Bohr's principle of complementarity (BPC) has been the cornerstone
of quantum theory since it was proposed in 1928
\cite{bohr1928,bohr1928nature}. This principle states that some
physical objects have multiple properties, but these properties are
exhibited depending on some types of exclusive detecting devices.
One well-known example is the wave-particle duality, by considering
a single particle in a two-way interferometer \cite{feynman1963}.
One can choose to observe the wave-like or particle-like behaviors
of the particle by using different detection arrangements.
Interference fringes have been observed for massive particles such
as neutrons \cite{summhammer1983}, electrons \cite{tonomura1989},
atoms \cite{carnal1991,keith1991} and molecules \cite{arndt1999},
all thought to be only particle-like before. These observations
shows the unfamiliar wave-like side of these particles. In the case
of light, both the anti-bunching effect and its interference
fringes--associated with its particle-like and wave-like properties
respective--have been previously demonstrated
\cite{grangier1986,braig2003,jacques2005}.

Besides these all-or-nothing situations, there actually exists some
intermediate stages
\cite{wootters1979,rauch1984,summhammer1987,buks1998,durr1998nature},
where the which-path knowledge corresponding to the particle-like
property is partially detected, resulting in the reduced
interference visibility. This issue was first discussed by Wooters
and Zurek in 1979 \cite{wootters1979}. Later, an inequality was
experimentally shown by Greenberger and Yasin in some unbalanced
neutron interferometry experiments
\cite{greenberger1988,mandel1991}, and theoretically derived by
Jaeger \emph{et al.} \cite{jaeger1995} and Englert
\cite{englert1996,englert2000} independently. This inequality is
written as
\begin{equation} \label{egdr}
V^{2}+D^{2}\leq 1,
\end{equation}
where $V$ is the visibility of the interference fringes, and $D$ is
the path distinguishability of the particle, which stands for the
available quantity of which-path knowledge from the system. This
inequality is also known as the EG duality relation. Plenty of
experiments have demonstrated this inequality with atoms
\cite{durr1998}, nuclear magnetic resonance
\cite{peng2003,peng2005}, faint laser \cite{schwindt1999}, and also
single photons in a delayed-choice scheme \cite{jacques2008}.
Recently, this duality relation has been extended to the more
general case of an asymmetric interferometer where only a single
output port is considered, and this inequality still holds
\cite{lili2012}.

One of the most efficient quantum systems for testing BPC is the
single photons in a Mach-Zehnder interferometer (MZI). In Ref.
\cite{jacques2008}, a series of unbalanced beam splitters (BS) were
randomly chosen in the MZI, including the extreme cases with
reflection coefficients of $R=0$ and $0.5$. However, we notice that
beam splitters of this type are all classical devices. Mapping to
the quantum BS (q-BS) scheme recently proposed by Ionicioiu and
Terno \cite{ionicioiu2011,Schirber2011}, the same results will come
out when the q-BS is selected to collapse on a set of eigenstates.
These eigenstates of the q-BS can be the same as the
previously-mentioned classical devices.

In our experiment, the q-BS stays at the quantum superposition
states of the extreme eigenstates--noted as $|a\rangle$ $(R=0)$ and
$|p\rangle$ $(R=0.5)$--corresponding to the absence and presence of
a balanced BS, respectively. We introduce this q-BS into the MZI,
and not only the eigenstates but also the quantum-superposition
states of the q-BS are selected as the bases to collapse on at
detection. The particles are single photons emitted from an
InAs/GaAs self-assembled quantum dot \cite{tangjs2009,licf2011}. Our
result shows that the EG duality relation is exceeded when some
certain detecting basis of the q-BS is chosen. This exceeding is
caused by the interference between the wave and particle properties
of the photons. In order to derive a generalized EG duality
relation, we consider both of the two orthogonal detecting bases,
then we find the generalized EG duality relation holds for our
results.

The experimental setup is sketched in Fig. 1(a). The single photons
are split by a $50:50$ BS into two paths, followed by a $\varphi$
phase, then combined by a q-BS. The use of the q-BS is the main
difference between this setup and a regular MZI. The photons are
detected by the single-photon avalanche photodiodes (APD).

\begin{figure}[tb]
\centering
\includegraphics[width=0.45\textwidth]{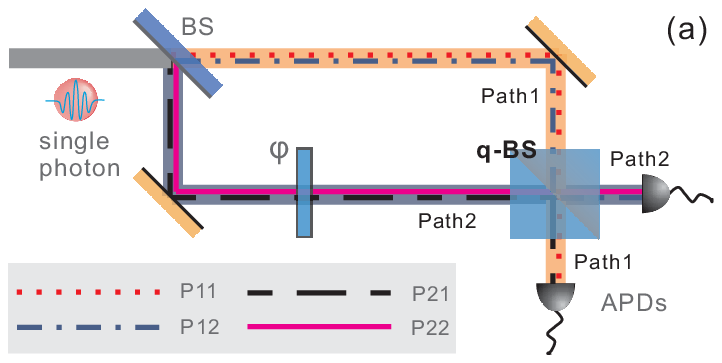}
\includegraphics[width=0.45\textwidth]{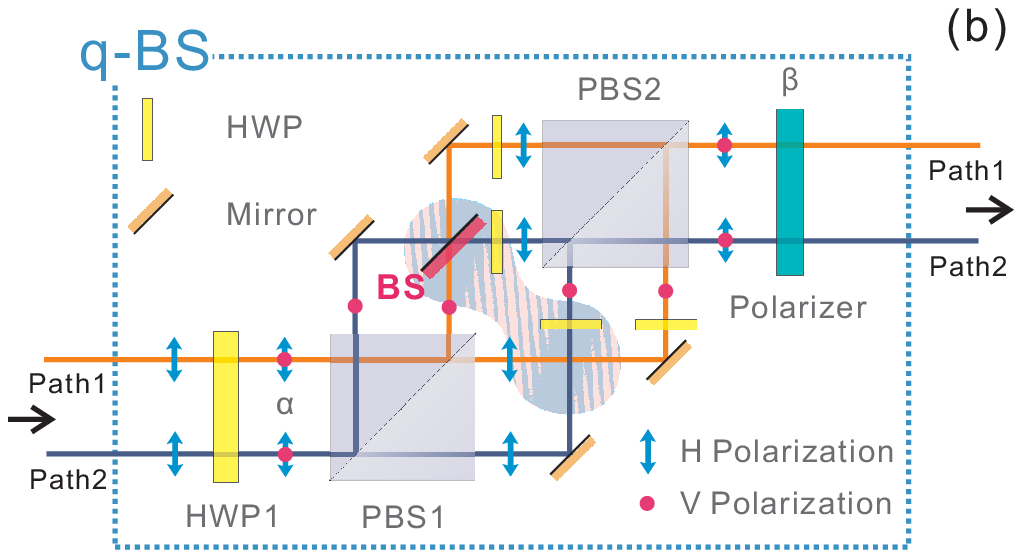}
\caption{\label{Fig1} (a) The MZI with a q-BS. The main difference
between this setup and a regular MZI is that the second BS is
replaced with a q-BS. Pij (i,j=1,2) are the four possible sub-paths
for the single photon used to define the distinguishability $D$. (b)
The simplified setup of the q-BS. Path 1 and Path 2 are both divided
into two components, which are in the quantum superposition states.
Each component corresponds to an eigenstate of the photon
polarization. One component constructs the closed MZI (a BS is
present) and the other constructs the open MZI (no BS). PBS2 then
recombines these two components, making a quantum-superposition
state of the closed and open MZIs. The direction of the photon
polarization before PBS1, $\alpha$, controls the states of q-BS. The
polarizer with a $\beta$ oriented axis selects the detecting basis
of the q-BS. }
\end{figure}

As discussed in Ref. \cite{lili2012}, we need to derive the photon
state after the q-BS and know the probabilities of each path taken
by the photon, in order to calculate the visibility. The state of
the q-BS is $|qbs\rangle=\frac{1}{\sqrt{2}}(|a\rangle+|p\rangle)$;
hence we derive the photon state (before the q-BS state is detected)
as
\begin{equation} \label{psistate}
|\psi\rangle=\frac{1}{\sqrt{2}}|particle\rangle|a\rangle+\frac{1}{\sqrt{2}}|wave\rangle|p\rangle
\end{equation}
according to Ionicioiu and Terno \cite{ionicioiu2011}, with
$|particle\rangle=\frac{1}{\sqrt{2}}(|1\rangle+e^{i\varphi}|2\rangle)$
corresponding to the particle state, and
$|wave\rangle=e^{i\frac{\varphi}{2}}(cos\frac{\varphi}{2}|1\rangle
e^{i\delta_{1}}-isin\frac{\varphi}{2}|2\rangle e^{i\delta_{2}})$
corresponding to the wave state. $\delta_{1}$ and $\delta_{2}$ are
two additional constant phases, which can be adjusted in the
experiment. The q-BS state is then collapsed on an arbitrary basis
$|b\rangle=sin\beta|a\rangle+cos\beta|p\rangle$, which means the
photon state becomes $\rho=\tilde{\rho}/Tr(\tilde{\rho})$, where
$\tilde{\rho}=Tr_{q-BS}(P_{b}|\psi\rangle\langle\psi|)$ with
$P_{b}=|b\rangle\langle b|$ as the projection operator. Here we
derive the probability that the photon takes Path 2 as
$p_{2}(\varphi)=Tr(|2\rangle\langle 2|\rho)$. From this probability,
we have the visibility of Path 2,
\begin{equation} \label{vis2}
V=\frac{p_{max}-p_{min}}{p_{max}+p_{min}}.
\end{equation}
As shown in Fig. 1(a), each photon has four possible sub-paths to
reach the APDs from the first BS (P11, P12, P21, P22). The photons
that finally appear on Path 2 can come from either P12 or P22, each
of which represents a totally different which-path knowledge.
Assuming that the probabilities of the photons coming from P12 and
P22 are respectively $w_{12}$ and $w_{22}$, the distinguishability
of Path 2 can be written as
\begin{equation} \label{dist2}
D=|w_{12}-w_{22}|.
\end{equation}
When the photons definitely come from either P12 or P22, then $D=1$;
when the chance that the photons are coming from either of the two
paths is equal, $D=0$. The same definitions of $V$ and $D$ are also
found in Ref. \cite{lili2012}, where the inequality (\ref{egdr}) is
proven to be correct for a general situation using the classical
unbalanced beam splitters.

For our experiment, the q-BS is realized by using the photon
polarization state as an ancilla to control the absence or presence
of the BS. The simplified setup of the q-BS is shown in Fig. 1(b).
Photon polarization for either path is first rotated by HWP1
(half-wave plate) in the direction of $\alpha$, which corresponds to
the q-BS state of
$|qbs\rangle=sin\alpha|a\rangle+cos\alpha|p\rangle$. For this
experiment, we fix this angle as $\alpha=45^{\circ}$. The photons
are then split by PBS1 (polarizing beam splitter) into two
components. In one direction, the photons go through a closed MZI
with a $50:50$ BS; in the other direction the photons go through an
open MZI with no BS. The two components are then recombined by PBS2,
with the photon state at that point exactly described by Eq.
(\ref{psistate}) with $|a\rangle\leftrightarrow|V\rangle$ and
$|p\rangle\leftrightarrow|H\rangle$. Note that $|H\rangle$ and
$|V\rangle$ represent the horizontal and vertical polarization
states of the photons, respectively. The polarizer set at the angle
of $\beta$ chooses the detecting basis of the q-BS.

We use the beam displacer (BD) actually to construct the MZIs
instead of the regular BS or PBS, i.e., the first BS in Fig. 1(a)
and the BS, PBS1 and PBS2 in Fig. 1(b). The BD moves the
extraordinary beam to a parallel path separated from the ordinary
beam by $4$ $mm$, making the MZI more stable than a traditional BS
system. Simplified sketches of the setup are given in Fig. 1, as
they illustrate the setup better than diagrams depicting all the
elements of the setup and aid in understanding. A more detailed
description of the setup can be found in our previous work
\cite{tang2012}.

\begin{figure}[tb]
\centering
\includegraphics[width=0.45\textwidth]{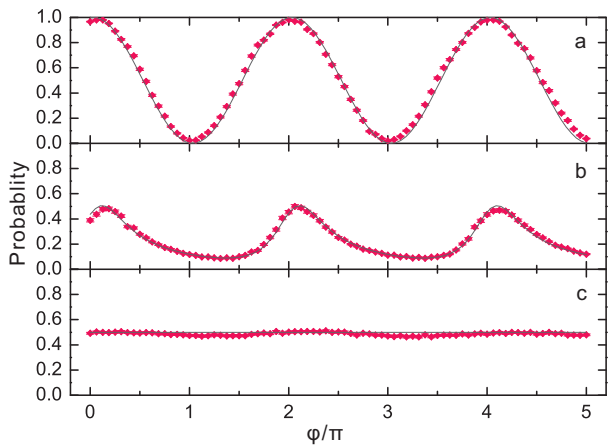}
\caption{\label{Fig2} Probability that the photon takes Path 2. (a),
(b) and (c) correspond to the cases of $\beta=0,$ $\frac{3\pi}{16}$
$and$ $\frac{\pi}{2}$, respectively. The solid lines are the
corresponding theoretical fits for each case.}
\end{figure}

To measure the visibility, we leave both paths in Fig. 1(a)
unblocked, count the photon numbers detected by the APDs, then
calculate the probability that the photon takes Path 2, i.e.,
$p_{2}(\varphi)$. The results are shown in Fig. 2. The solid lines
are the theoretical fits corresponding to each set of experimental
data. Fig. 2(a) is the $\beta=0$ case, where the q-BS state is
detected on the basis of $|b\rangle=|p\rangle$, which is the
eigenstate associated with the closed MZI. Therefore, the photons
behave as a wave, and the visibility (shown in Fig. 3(a)) of the
interference fringe reaches $0.961\pm 0.004$. This result coincides
with the classical-BS-experiment result found in Ref.
\cite{jacques2008}. Fig. 2(c) corresponds to the
$\beta=\frac{\pi}{2}$ case. Similarly, the q-BS state is detected on
the other eigenstate $|b\rangle=|a\rangle$, which is associated with
the open MZI. Thus, the photons behave as particles. The result is
also the same as in the classical BS case. However,
$\beta=\frac{3\pi}{16}$ for Fig. 2(b), so the detecting basis here
is a quantum-superposition state, which is related to the MZI
staying in both a closed and an opened state. The visibility in this
case is $0.707\pm 0.017$. The photons behave as a quantum
superposition of wave and particle, which is well illustrated by the
expression describing the photons' state $\rho$, i.e.,
$C_{1}(sin\beta|particle\rangle+cos\beta|wave\rangle)$ (where
$C_{1}$ is a coefficient). This phenomenon does not have a
counterpart in the classical BS experiment. The differences between
the experimental and theoretical values are caused by the counting
statistics, the imperfection of the optical glasses, the dark and
background counts, and the tiny instability of the MZIs.

\begin{figure}[tb]
\centering
\includegraphics[width=0.45\textwidth]{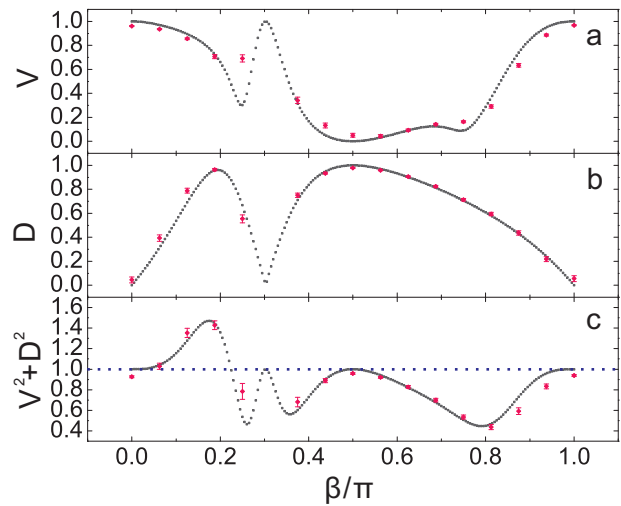}
\caption{\label{Fig3} (a) The visibility $V$, (b) The path
distinguishability $D$, and (c) $V^{2}+D^{2}$. The dashed line in
(c) is the limit of the EG duality relation ($1$), which is exceeded
in this situation.}
\end{figure}

To measure the distinguishability $D$, we first block Path 1 after
the BS in Fig. 1(a) and detect the number of photons coming from P22
($N_{22}$), then block Path 2, and detect the photon number from P12
($N_{12}$). Hence, the distinguishability of Path 2 can be
calculated using $D=\frac{|N_{12}-N_{22}|}{N_{12}+N_{22}}$ according
to Eq. (\ref{dist2}). The result is shown in Fig. 3(b) with larger
dots, and the smaller-dot line is the theoretical simulation. When
$\beta=0$ (the closed MZI), then $D=0.045\pm 0.024$ and no
which-path knowledge is available. However, when
$\beta=\frac{\pi}{2}$ (the open MZI), then $D=0.97751\pm 0.0038$ and
full which-path knowledge is detected. This result is in accord with
the wave-like and particle-like behavior of the photons previously
discussed. For these all-or-nothing cases, the q-BS collapses on the
eigenstates, which means these situations give the same results as
the classical BS experiment; the inequality (\ref{egdr}) holds, and
the upper bound is reached (See in Fig. 3(c)). On the other hand, in
the quantum intermediate case of $\beta=\frac{3\pi}{16}$, the value
of $V^{2}+D^{2}$ goes beyond the limit of the EG duality relation
($1$, the blue dashed line in Fig. 3(c)) by $10$ deviations to
reache $1.428\pm 0.043$. This result coincides with the results from
the theoretical simulation.

This exceeding of the EG duality relation is caused by the quantum
superposition of the photons' wave and particle states--or the
interference between them--introduced by the q-BS and a quantum
intermediate detecting basis. To illustrate this point and derive a
generalized EG duality relation, we combine the corresponding photon
counts of the two orthogonal bases related to $\beta$ and
$\beta+\frac{\pi}{2}$, then calculate $V_{g}^{2}+D_{g}^{2}$ in the
same way. The forms of $V_{g}$ and $D_{g}$ are the same as $V$ and
$D$, respectively. However, the photon counts and the meanings are
different. The former ones correspond to the sum of the counts of
two orthogonal bases, and describe the behavior of photons in these
two cases as a whole; the wave-particle interference becomes an
internal effect here. On the other hand, the later ones describe the
behavior of photons in a single basis case. We find that the
generalized inequality ($V_{g}^{2}+D_{g}^{2}\leq 1$) holds for our
results, shown in Fig. 4(a). The solid line is the theoretical
simulation. To further analyze this combination process, we
calculate the final state of the photon after the combination, found
to be $C_{2}(sin^{2}\alpha|particle\rangle\langle
particle|+cos^{2}\alpha|wave\rangle\langle wave|)$, where $C_{2}$ is
a coefficient. This state is a classical mixture of the wave and
particle properties, and is independent of the chosen orthogonal
basis pair (defined by $\beta$). However, the state is related to
the parameter $\alpha$, which determines the state of the q-BS and
also the probabilities of the photon going through the closed or
open MZIs. $V_{g}^{2}+D_{g}^{2}$ is calculated to be
$sin^{4}\alpha+cos^{4}\alpha$, which is not larger than 1; when
$\alpha=\frac{\pi}{4}$, then $V_{g}^{2}+D_{g}^{2}=0.5$. We have also
measured $V_{g}^{2}+D_{g}^{2}$ using various values of $\alpha$,
with the result shown in Fig. 4(b), which further proves our
previous discussions. There is a systematic error in Fig. 4, which
may caused by the dark and background counts, the decoherence
processes, the imperfection of optical glasses and the imprecision
of experimental parameters.

\begin{figure}[tb]
\centering
\includegraphics[width=0.45\textwidth]{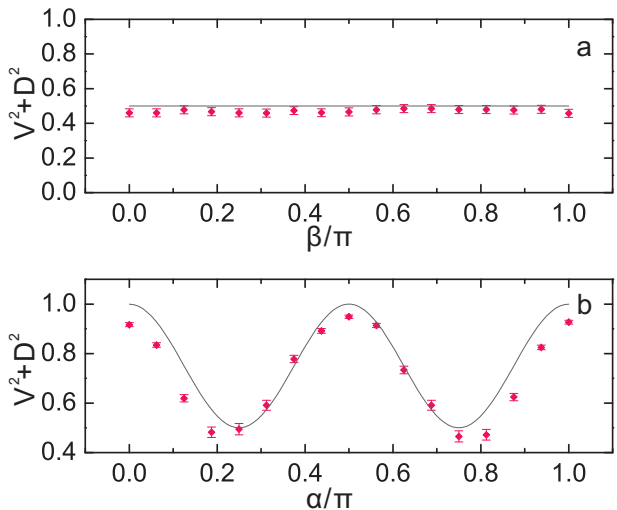}
\caption{\label{Fig4} $V_{g}^{2}+D_{g}^{2}$ after combination of the
photon numbers of two orthogonal-basis cases with (a) varying
$\beta$ and fixed $\alpha=\frac{\pi}{4}$ and (b) varying $\alpha$
and arbitrary $\beta$. The generalized EG duality relation holds for
these results.}
\end{figure}

Actually, the violation of BPC--and specifically the EG duality
relation--has been declared by Afshar \emph{et al.}
\cite{afshar2007}, who believe that quantum mechanics is not
correct, but others disagree with this interpretation
\cite{steuernagel2007,jacques2008njp,georgiev2007,georgiev2012}, and
the debate continues. We must note here that our experiment is
completely unrelated to the Afshar experiment. Even though our
results exceed the EG duality relation, our experiment as a whole is
in accord with quantum theory and is only a small extension of BPC,
i.e., the classical detecting devices are replaced with the quantum
devices for our experiment. In the original BPC, the detecting
devices can only be in the classical states, which are each related
to the properties that can be shown. In contrast, the detecting
devices can exist in the quantum-superposition states in our
extension by using the quantum control \cite{ionicioiu2011}. This
small change makes the originally exclusive properties of the object
appears to be quantum-superposed, allowing for the limit of the EG
relation duality to be exceeded.

Besides experiments in wave-particle duality, there are many other
well known experiments whose results form the foundation of quantum
mechanics: the Bell-inequality experiments
\cite{Bell1964,Clauser1969,Freedman1972}, the
Kochen-Specker-inequality experiments
\cite{Kochen1967,Mermin1990,huangyf2003}, and so on. The new concept
of using a quantum device could also be introduced into these
experiments, potentially allowing new phenomena to appear, which
could further our understanding of quantum mechanics.

In conclusion, we introduce a q-BS, proposed in Ref.
\cite{ionicioiu2011}, into the unbalanced MZI used in Ref.
\cite{jacques2008}, selecting some quantum-superposition states of
the q-BS as the collapsing bases to detect the q-BS's states.
Following the definitions of visibility and distinguishability used
in Ref. \cite{lili2012}, we find the limit of the EG duality
relation is exceeded. We conclude that this result is caused by the
interference between the wave and particle properties of the
photons. After we combine the corresponding photon numbers of two
mutually orthogonal collapsing bases of q-BS, the wave-particle
interference becomes an internal effect, then the generalized EG
duality relation holds. This work is entirely within standard
quantum theory, but opens up a new way for people to understand the
quantum world by replacing the classical devices with quantum ones.

This work is supported by the National Fundamental Research Program,
National Natural Science Foundation of China (Grant No. 60921091,
10874162).




\end{document}